\documentclass[11pt]{article}
\usepackage{moriond,epsfig}
\def\be{\begin{equation}}
\def\ee{\end{equation}}
\def\bea{\begin{eqnarray}}
\def\eea{\end{eqnarray}}

\begin{document}
\vspace*{4cm}
\title{MULTI(SCALE)GRAVITY: A TELESCOPE FOR  THE MICRO-WORLD}

\author{ Ian I. Kogan }

\address{Department of Physics, Theoretical Physics, 1 Keble Road,\\
Oxford OX1 3NP, England}

\maketitle\abstracts{
This is a talk presented at XXXVIth Rencontres de Moriond,
 ElectroWeak Interactions and Unified Theories, March 2001.
  A short review of modern status of multigravity, 
i.e. modification of gravity  at both short and large distances  is
given.}

\section{Introduction}
The Brane Universe scenario  is actually  quite an old idea 
\cite{Regge:1975iu,Akama:1982jy,Rubakov:1983bb,Visser:1985qm,Squires:1986aq,Gibbons:1987wg}.
\footnote{The author learnt  about Regge and Teitelboim  (RT) paper 
\cite{Regge:1975iu} which is virtually unknown in a modern ``Brane 
community'' 
from \cite{Cordero:2001qd}. For  more information about RT see 
\cite{Davidson:1999ys} and references therein; see also 
\cite{Deser:1976qt} for critical comparison between RT approach and 
four-dimensional general relativity. In \cite{Gibbons:1987wg} there
are also references on even earlier discussions on  what is called now
 ``Brane Universe'' - but was called ``Embedding Problem'' in 1965. 
Proceedings of seminar on embedding problems edited by I.Robinson and 
Y.Ne'eman are published in \cite{embedding}.  It seems that
the more we work on Brane Worlds the more  ideas from the past
become visible (the same way as light from distant stars we could not
see earlier). Perhaps when we  reach some future frontier an
old picture about Earth (read ``Universe'') standing on three
elephants standing on one big turtle will make some sense ?}
Recently it has been subject of renewed interest 
\cite{Arkani-Hamed:1998rs,Antoniadis:1998ig,Arkani-Hamed:1999nn} with
 the realization 
that such objects are common in string theory. In particular, there has  
been  a lot of activity  on warped brane  constructions in five 
spacetime 
dimensions, motivated by heterotic M-theory 
\cite{Horava:1996qa,Witten:1996mz,Horava:1996ma} and its five 
dimensional reduction \cite{Lukas:1999yy,Lukas:1999qs}. In the context  
of these constructions, one can localize gravity on the brane world 
having four dimensional gravity even with an extra dimension of 
infinite extent \cite{Gogberashvili:1999tb,Randall:1999vf}, or can 
generate an exponential mass hierarchy on a compact two brane model as 
it 
was done in the Randall - Sundrum (RS1) model \cite{Randall:1999ee},  
providing a novel geometrical resolution of the Planck hierarchy 
problem. For more details on warped 
models see an excellent review of \cite{Rubakov:2001kp}. 
 Usually embedding of Standard Model and General Relativity into any
multidimensional construction gives rise to all possible sorts of new
effects in a Micro-world, i.e. we are definitely going to modify a
short-distance physics. 
The subject of this talk is to show that  sometimes one can get a very 
drastic modification of the laws of gravity at ultra-large,
cosmological scale.

\section{Modification of gravity at  large distances}
The force of gravity between two bodies with masses $m_1$ and $m_2$ 
depends on a distance $r$  between them as 
\bea
F = G \frac{m_1 m_2}{r^2}
\eea
This law was used by Newton to explain  the famous 
Kepler laws\footnote{
Actually there are strong evidence (see for example very interesting
book  \cite{parsimon} and references therein)
that Robert Hooke  suggested the inverse square law to Newton. Hooke
 used the third Kepler's law $R^3 \omega^2 = const$ and expression for
centrifugal force  $\omega^2 R$ 
acting on a particle moving with a constant angular velocity $\omega$ 
along a circle of radius $R$ to get the force proportional to $1/R^2$.
 It was also known to Halley and Wren. But only Newton was able to
derive all three Kepler's laws. As another curious element of Oxford
involvement in gravity research in the 17 century it is interesting to 
note that  first meetings of Royal Society were held in Wadham
College, Oxford }. 

Why it is inverse square law ? The modern answer is the following -
 because we live in a  three-dimensional space (four-dimensional
space-time) and graviton is massless !

How well do we know this law ? We definitely do not know it below 
$1-0.1 mm$ as well we do not know it at distances comparable with 
 the size of observable Universe.  
Can we modify it at large scales ? \footnote{In this talk we are not
going to discuss the modification of gravity  below $1 mm$ - 
short-distance modifications are inevitable in any
quantum/multidimensional/etc scenario. One can get it for example
 in a pure four-dimensional theory with $R^2$ terms - see 
\cite{Tekin:2001xj}
 and references therein.} There are a lot of publications about
possible modification of Newtonian gravity at large  scales, for
example famous Milgrom  proposal \cite{Milgrom:1998aj,Milgrom:1998yb}
as well as many others - for the list of references see
\cite{Aguirre:2001xs}. The problem with these modifications that it is 
hard to reconcile them with General Relativity. 

 In General Relativity the deviation from the inverse square law must
be due to a very small mass of a graviton.
 However it seems that there
 is no consistent four-dimensional theory 
with massive graviton \cite{Boulware:1972my} \footnote{For early
discussion of phenomenological limits on graviton mass see \cite{Goldhaber:1974wg}.}

\subsection{Giant  see-saw  with us in the middle}

Before we shall discuss multidimensional theories let us ask a 
 question - what is the largest see-saw  one can make ?
 Well, in physics see-saw means that there are 3 scales -
 large, small and  the third one whis is just the geometric average
 of the first two. Than the largest possible see-saw is a such one
that the large scale is the largest possible scale and the second one
- is the smallest possible scale.  The largest scale we have now
is  the size of observable Universe $R_{U} = 10^{28}$ cm and the
smallest one must be the Planck length $r_P = 10^{-33}$ cm. 
 So what will be the third, intermediate scale $L$ ?
\bea
r_P \,R_{U} = L^2, \,\,\,\, L^2 = 10^{-5} cm^2, \,\,\,\, L \sim
10^{-1}-10^{-2} mm
\label{eq:seesaw}
\eea
It is really amusing that in the middle of the largest see  we
 can find a place for ourselves. This scale is a ``biological''  one
 -  close to the size  of our cells, etc...
 
But of of course this  new scale $L$  can be related to some  non
biological
 physics too.
 First of all it corresponds to mass scale $10^{-3}-10^{-4}$ ev
which is very close to the current limit on cosmological
constant\footnote{For recent review - see \cite{Carroll:2001fy}.} 
(neutrino physics also may be related to this energy scale). The reason
for this 
 is very simple - the curvature radius $R$ corresponding for 
 the cosmological constant $\Lambda$ is given by $R^{-2} = \Lambda 
r^2_P$.
 If vacuum energy is $1/L$ then $\Lambda \sim  (1/L)^4$ and we get 
 $R^{-1} \sim r_P/L^2$. The current limit for cosmological constant
  is such that respective  curvature radius $R \sim R_{U}$ - so we
 again have  Eq.~\ref{eq:seesaw}.

Another  interesting fact about  the scale $L$ 
 is that this is just the boundary between  quantum 
and classical behaviour for particle with a Planck  mass scale. If one
  considers the wave packet of intimal size $L$ for  particle with mass
 $m$   the  wave packet size at moment $t$ will be (we use units
 $c= \hbar = 1$)
 \bea
L^2(t) = L^2 + \frac{t^2}{m^2L^2} 
\eea
from which one can see that the transition time $T$ between classical 
and quantum behaviour is given by $T \sim m L^2$. For Planck mass
particle $m = r^{-1}_{P}$  the  scale $L\sim
10^{-1}-10^{-2} mm$ corresponds to transition
 time equal to the age of Universe $T \sim R_{U}$ as we can see from 
Eq.~\ref{eq:seesaw}.

 As we shall see later the giant see-saw means something else -
 modification of Newtonian gravity at large scales  can not be excluded
 \cite{Kogan:2000wc}. But to see how does Eq.~\ref{eq:seesaw} lead to 
such
 a  surprising possibility we have to  discuss first how does one get
four-dimensional gravity from a multi-dimensional gravity.
 
\subsection{Four-dimensional gravity from extra dimensions}
Let us start from a simple question - how one can get a 
four-dimensional
gravity
 from  higher dimensions. The Einstein-Hilbert action in D-dimensional
space-time  is
\bea
S = M_{F}^{D-2} \int d^{D-4}y d^{4} x \sqrt{G} G^{MN} R(G)_{MN}
\eea
where $M_{F}$ is some fundamental scale and $G_{MN}$ is a full
D-dimensional
 metric describing in a flat space limit graviton with $D(D-3)/2$ 
degrees
of freedom.
 For a four-dimensional  gravity to exist  we must have  finite Planck
mass
 $M_P$ which is defined as 
\bea
M_{P}^{2} = M_{F}^{D-2}\int d^{D-4}y \sqrt{G}
\eea
This existence of a four-dimensional gravity  depends on what is the
volume of extra dimensions. If it is finite (even if the space is
non-compact) we bound to have  four-dimensional gravity. If it is 
infinite
- massless graviton  will not exist (but we may have a massive one).

But of course besides massless graviton we shall have massive 
excitations.
 For example in a canonical KK case of a space $R^4 \times S^1$  there 
is
a 
 a whole massive KK tower. Each massive graviton has 5 degrees of 
freedom
- precisely  the same number as massless graviton in five dimensions $ 
5
(5-3)/2 = 5$. Massless  graviton has only 2, but then there is a 
massless
vector field - graviphoton $G_{\mu 5}$ with 2 degrees of freedom and
scalar $G_{55}$ - another
 degree of freedom. So at each muss level we have $5$ degrees of 
freedom
as it must be - compactification can not change the total number of the
degrees of freedom. 

So why we can not use these massive gravitons to modify gravity ? We 
can -
but this will be modification at {\it short} distances.  The mass 
spectrum
is equidistant  and for a circle with radius $R$ is given by  $m_{n} =
n/R$.
 We just can not take $m_1$ to be to light - the moment the first mode 
is
important we start to see the whole KK tower - and we open the new
dimension.
 The same is true for all known compact manifolds - they all have 
regular
spectra.  There is no way  to  modify gravity at large distances unless 
we
have
 something like our giant see-saw -  which one can write in another 
form
\bea
M_P m_1 = m_2^2
\label{eq:massseesaw}
\eea
where $M_P$ is a Planck mass, $m_1$ is an ultra-light mass of first (or
several first) excitation(s) corresponding to a cosmological scale, 
i.e.
$m_1 \sim 
R_{U}^{-1}$, where $m_2$ corresponds to much higher mass scale
corresponding
 to much smaller  length  $L$  below  sub-millimeter scale.
If there are no graviton masses between $m_1$ and $m_2$ we do not
 have any obvious violation of a Newton law between $L$ and $R_U$
- what we see is determined by massless and ultra-light graviton 
and  other excitations become important only at short distances 
 smaller than $L$.  Of course we can take 
 $m_1$ a little bit bigger and without any contradiction with
 experiment modify gravity at let say $1\%$ of a Universe size 
\cite{Kogan:2000wc}.  We can do other things (some of which will 
 be discussed later) - but independently on what model we are
 going to deal with there is one important general rule:
 we must have a spectrum with extremely light (ultra-light)
 states.
 
\subsection{Multigravity - ultra-light states from
multilocalization}\label{subsec:multi}

We must have some natural mechanism producing these ultra-light 
states - and as it always happens in physics we want  this mechanism to 
be natural. It means that in some limit our ultra-light states must 
become
 massless.  But then it is clear that we have to start from some
configuration
 which support massless graviton (and we  have already discussed that
there are many  ways to do it) and  after that do something unusual - 
take
several configurations like that and let them "talk to each other".
>From the basic principles of quantum mechanics we know that several
degenerate levels will split - and  if the system is stable and there 
are
no tachyons, we shall get  a  massless ground state and all other 
states become ultra-light. 
This is true for compact spaces - noncompact spaces can be  studied as
limiting cases of compact ones.

 So the non-trivial task is to figure out how is it possible to 
 take several configurations  supporting massless graviton
 and let them "talk to each other" ? The only solution known now is 
based on 
  multibrane constructions.  There is no place here to discuss
 technical details so we shall  explain only general concepts, for more
 details reader  is advised to look at original papers.

It is obvious that in order the scenario of multi-localization  to be
realized it is necessary to have a mechanism that induces the
localization of the fields under consideration. For the case of the
graviton, it has been shown that it can be localized if the geometry
of the extra dimension is non trivial (with specific 
properties).Remember
that  what we  need  is  more  than  simply a massless graviton (in 
which
case   finite
 volume of extra dimensions will be enough) - but massless graviton  
which
is
{\it localized}. As an illustration let's consider a 
non-factorizable geometry with one extra dimension. In this scenario
the fifth dimension $y$ is compactified on an orbifold, $S^{1}/Z_{2}$
of radius $R$, with $-L \le y \le L $. The five dimensional
spacetime is a slice of $AdS_{5}$ which is described by\footnote{We
will assume that the background metric is not modified by the presence 
of the bulk fermion, that is, we will neglect the back-reaction on the 
metric from the bulk fields.}: 
\be
ds^{2}=e^{-2 \sigma(y)} \eta_{\mu \nu}dx^{\mu}dx^{\nu}+dy^2
\ee
where the warp factor $\sigma(y)$  depends on the details of the model 
considered. For the moment we assume that
we have a model with a number of positive and negative tension flat
branes (the sum of the brane tensions should be zero if one wants 
flat four dimensional space on the branes) and that this function is 
known ( it can be found by
looking the system gravitationally). 

Starting from a five dimensional
Lagrangian in order to give four dimensional interpretation to the
five dimensional fields one has to go through the dimensional
reduction procedure. This procedure includes    
the decomposition of the  five dimensional fields  (actually these
 arguments can be applied not only to gravity and this is why we 
  do not specify the  tensor structure)
$\Phi(x,y)$ in  KK states:
\bea
\Phi(x,y)=\sum_{n=0}^{\infty} \Phi^{(n)}(x) f^{(n)}(y)
\eea
where $f^{(n)}(y)$ is a complete orthonormal basis. The idea behind
this KK decomposition is to find an equivalent description of the five
dimensional physics associated with the field of interest, through an
infinite number of KK states with mass spectrum and couplings that
encode  all the information about the five dimensions.
The function $f^{(n)}(y)$ describes the localization of the
wave function of the n-th KK mode along the extra dimension. In order
the above  to be possible, it can be
shown that 
 $f^{(n)}(y)$ should obey a second order differential
equation which after a convenient change of variables and/or a
redefinition\footnote{The form of the redefinitions depend on the spin 
of the field.} of
the wave-function, reduces to an ordinary
Schr\"{o}dinger equation:
\bea
\left\{-\frac{1}{2}
\partial_z^2+V(z)\right\}\hat{f}^{(n)}(z)=
\frac{m_n^2}{2}\hat{f}^{(n)}(z)
\eea
The mass spectrum and the wave-functions (and thus the couplings)
are determined by solving the above differential equation. Obviously
all the information about five dimensional physics is contained in the
form of the potential $V(z)$. For example in the case of the graviton
the positive tension branes correspond to attractive $\delta$-function 
potential wells
whereas negative tension branes to  $\delta$-function barriers. The
form of the potential between the branes is determined by the
$AdS_{5}$ background.

 \subsection{Example: +--+ model as interpolation between
 Bigravity and quazilocalization}
 The first model where  multigravity been found is the 
so-called $+-+$  or bigravity \footnote{sometimes also called 
``Millennium''
 being written actually a year before the (correct) new millennium}
 model   \cite{Kogan:2000wc}, which consists 
of two positive  branes with tensions $\Lambda$ located 
at the fixed points of a $S_1/Z_2$ orbifold with one negative 
brane with tension $-\Lambda$ which can move freely in
 between. The next one is a GRS model \cite{Gregory:2000jc} which 
can be obtained from this model by cutting the negative
brane in half, i.e. instead of one  $''-''$ brane with tension 
$-\Lambda$
 one can take two branes with negative tension $-\Lambda/2$ each and
then move them apart. Generic case was suggested  and studied in 
\cite{Kogan:2000cv,Kogan:2000xc} and is called $+--+$ model (see
Fig.1)

\begin{figure}[ht]
\begin{center}
\mbox{\epsfig{file=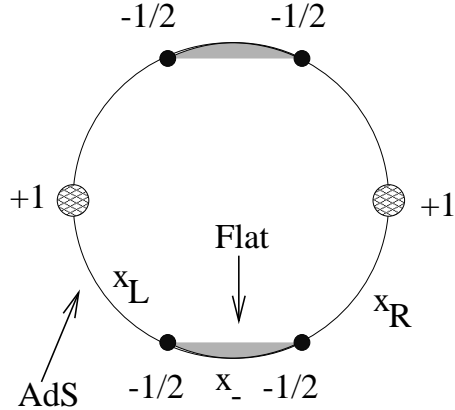,width=6cm}}
\caption{$+--$ model with two $''+''$ branes at the fixed points and 
 two moving $''-1/2''$ branes. 
In the limiting case when $x_{-} \rightarrow 0$ we have a $+-+$ model. 
The GRS model can be obtained in the opposite limit 
 $x_{-} \rightarrow \infty$
 The whole configuration can be
 considered as  two GRS models connected via flat space.
\label{Fig.1}}
\end{center}
\end{figure}
This model  consists of four parallel 3-branes in an $AdS_5$ space with
cosmological constant $\Lambda<0$. The 5-th dimension has the geometry
of an orbifold and the branes are located at
$y_0=0$, $y_1 = l_L$, ~  $y_2 = l_L+ l_{--}$ and $y_3 =y_2 + l_R$,
 where $y_0$ and $y_3$
are the orbifold  fixed
points\footnote{The requirement that we have orbifold fixed points is
not really necessary for our analysis, which is much more general}
 
Firstly we consider the branes having no  matter on them in
order to find a suitable vacuum solution. The action of this setup is:
\bea
S=\int d^4 x \int_{-y_3}^{y_3} dy \sqrt{-G} 
\{-\Lambda + 2 M^3 R\}-\sum_{i}\int_{y=y_i}d^4xV_i\sqrt{-\hat{G}^{(i)}}
\eea
where $\hat{G}^{(i)}_{\mu\nu}$ is the induced metric on the branes
and $V_i$ their tensions. Here we have   included  negative $y$ and 
we look for solutions invariant with respect to $Z_2$ symmetry 
$y\rightarrow -y$.

At this point we demand that our metric respects 4D Poincar\'{e}
invariance. The metric ansatz with this property is the following:
$
ds^2=e^{-2\sigma(y)}\eta_{\mu\nu}dx^\mu dx^\nu +dy^2
$
Here the ``warp'' function $\sigma(y)$ is essentially a conformal
factor that rescales the 4D component of the metric. 
It satisfies the following differential equations:
$
\left(\sigma '\right)^2=k^2, ~~
\sigma ''= \sum_{i}\frac{V_i}{12M^3}\delta(y-L_i)
$
where $
k=\sqrt{\frac{-\Lambda}{24M^3}}$ is a measure of the curvature of the 
bulk. 
The brane tensions are tuned to $V_0=-\Lambda/k>0$,
$V_1= V_2 = \Lambda/2k<0$, \mbox{$V_3=-\Lambda/k>0$}. 
It is convenient to introduce 3  dimensionless parameters
$x_L = kl_L, ~~~ x_R = k l_R, ~~~ x_{-} =   k l_{--}
$

We consider now the spectrum that follows from  dimensional reduction.
 This requires we find the spectrum of  (linear) fluctuations of the 
metric:
\bea
ds^2=\left[e^{-2\sigma(y)}\eta_{\mu\nu} +\frac{2}{M^{3/2}}h_{\mu\nu}
(x,y)\right]dx^\mu dx^\nu +dy^2
\eea

We expand the field $h_{\mu\nu}(x,y)$ in terms of the graviton and KK 
plane wave states :
$
h_{\mu\nu}(x,y)=\sum_{n=0}^{\infty}h_{\mu\nu}^{(n)}(x)\Psi^{(n)}(y)
$
where
$\left(\partial_\kappa\partial^\kappa-m_n^2\right)h_{\mu\nu}^{(n)}=0$
and the gauge is fixed to satisfy 
$\partial^{\alpha}h_{\alpha\beta}^{(n)}=h_{\phantom{-}\alpha}^{(n)\alpha}=0$.
The function $\Psi^{(n)}(y)$ obeys a second order differential
equation which, after a change of variables, reduces to an ordinary
Schr\"{o}dinger equation:
\bea
\left\{-
\frac{1}{2}\partial_y^2+V(y)\right\}\hat{\Psi}^{(n)}(y)=
\frac{m_n^2}{2}\hat{\Psi}^{(n)}(y), ~~~~
\hat{\Psi}^{(n)}(y)\equiv \Psi^{(n)}(y)e^{\sigma/2}
\eea
where the potential $V(y)$  is determined by $\sigma(y)$. Qualitatively 
it is made up of $\delta$-function potentials (attractive for $''+''$ 
and 
repulsive for $''-''$ branes) of
different weight depending on the brane tension plus a  smoothing
term  (due to the AdS geometry) that gives the attractive 
potentials a  ``volcano'' form. 

An interesting characteristic of this potential is that it always
gives rise to a (massless) zero mode which reflects the fact that
Lorenz invariance is preserved in 4D spacetime.  This mode is
normalizable for finite $x_{-}$ and becomes non-normalizable in the GSR 
limit  of infinite $x_{-}$.
The interaction of the linearized gravitons  with matter localized on a 
brane located at some  $y$ is given by 
$
{\mathcal{L}}_{int}=\frac{f(y)}{M^{3/2}}\sum_{n\geq 0}
\Psi^{(n)}(y)h_{\mu\nu}^{(n)}(x)T_{\mu\nu}(x) 
$
with $T_{\mu\nu}$ the energy momentum tensor of the SM Lagrangian and
 $f(y)$  some universal function. From this expression the Newton
potential  on a brane is given by
\be
U(r) \sim  \sum_{n\geq 0} \frac{(\Psi^{(n)})^2(y)}{M^3} 
\frac{e^{-mr}}{r}
~ \sim \int dm  \frac{\Psi_{m}^2(y)}{M^3} \frac{e^{-mr}}{r}
\sim  \int dm \rho(m)  \frac{e^{-mr}}{r}
\ee
where the  spectral density $\rho(m)$ is determined by the values of 
normalized wave functions $\Psi_{m}$. It  is discrete for $x_{-}=0$
 and any finite $x_{-}$ and becomes continuous in the GRS limit of 
infinite $x_{-}$.In the case $x_ =0 $  the ultra-light  mass  equals to  
$
m_1=2ke^{-x_L-x_R} 
$
which in symmetric case $x_L= x_R$  gives $m_1=2ke^{-2x}$. In this case 
the two  wave functions on the $+$ branes are equal,
  $ \Psi_{0}^2(0) = \Psi_{1}^2(0)$,  to  a high accuracy. The 
masses of the other KK states  are found to
depend in a different way on the parameter $x$. The mass of the
second state and the spacing $\Delta m$ between the subsequent states 
have the 
form:
$
m_2 \approx k e^{-x}, ~~~
\Delta m \approx \varepsilon k e^{-x}
$
where $\varepsilon$ is a number between 1 and 2. One can see
 that we just got our Giant see-saw!
For  length scales less than $m_1^{-1}$, gravity is generated by the 
exchange of {\it 
both} the massless graviton and the first KK mode, giving the
gravitational potential  
\be
U(r) = C \frac{(\Psi^{(0)})^2(0)}{M^3} \left(\frac{1}{r} +
  \frac{(\Psi^{(1)})^2(0)}{(\Psi^{(0)})^2(0)} \frac{e^{-m_1 
r}}{r}\right) +
 O(e^{-m_2 r})
 \approx  2C \frac{(\Psi^{0})^2(0)}{M^3} \frac{1}{r}
\ee
 where C is some constant. We see that  the gravitational constant is
 $G_N =  2C \frac{(\Psi^{0})^2(0)}{M^3}$.
According to this picture deviations from Newton's law will appear
in the submillimeter regime $ m_2 r < 1$ as the 
Yukawa corrections of the second and higher KK
states become important.  Also the
presence of the ultra-light first KK state will give deviations
from Newton's law as we probe cosmological scales $m_1 r >1$ 
(of the order of the observable universe) with $G_N/2$ instead of
$G_N$. The phenomenological signature 
of this scenario is that gravitational
interactions will appear to become weaker 
 for distances larger than $1/m_1 $.
 Of course when we have asymmetric case the absolute values of
 the wave functions on positive branes are not equal and one may have 
  arbitrary mixture of massless and ultra-light gravitons contribution
 to the gravitational constant.
 \begin{figure}[ht]
\begin{center}
\mbox{\epsfig{file=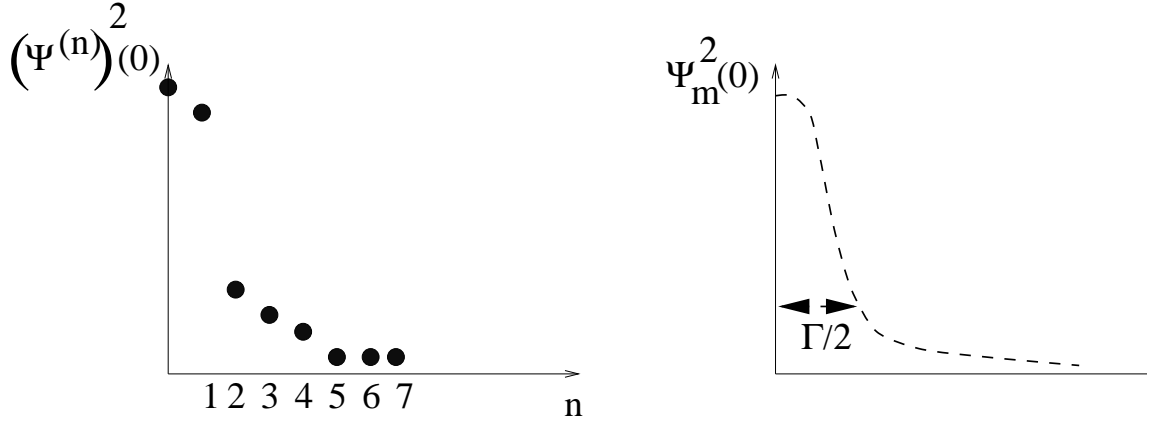,width=16cm}}
\caption{Behaviour $\Psi_{m}^2(0)$ in a $+-+$ model (discrete
spectrum). For a general $+--+$
model the discrete spectrum  density increases with the increase of
$x_{-}$ and spectrum becomes continuous in  GRS limit.
\label{Fig.2}}
\end{center}
\end{figure}

The  structure in the GRS construction is  totally different as they do 
not have normalized modes, but rather a continuous spectrum. 
 In this case there is a ``resonance'' effect  \cite{Csaki:2000pp}
 which is due to the fact that the negative brane creates a 
tunneling factor (the negative brane
acts as a repulsive potential)  which  effectively leads to  a
resonance in a wave function $\Psi_{m}(0)$ describing the  gravitons
$
\Psi_{m}^2(0) = \frac{c}{m^2 + \Gamma^2/4} + O(m^4)
\label{resonance}
$
It can be shown that for $r > 1/\Gamma$  the  potential 
is $1/r^2$ and gravity  is modified at large distances.
 But  instead of our Giant see-saw we have something else. 
 Let us  remind that in multigravity scenario there  are  three scales:
the Planck mass $M_P = r_P^{-1}$, the ultra-light mass scale $m_1$
(in GRS limit it is the width of resonance) which determines
 the  multigravity crossover scale $R_{C} = m_1^{-1}$ (cosmological scale
!)
 and  the next mass scale
 $m_2$ which defines the scale at which gravity is modified at short
 distances $L = m_2^{-1}$. In GRS limit the relation between them is
\bea
M_P^2 m_1 = m_2^3, \,\,\,\,\,\, R_C = L \left(\frac{L}{r_{P}}\right)^2
\label{eq:GRSseesaw}
\eea
which is different from Eq.~\ref{eq:massseesaw}. If now we shall take
 $L$ the same as before the crossover radius $R_C$ will be much bigger
 than horizon. To get multigravity we must have $m_2$ in a MeV range.
 Generic $+--+$ model gives us the following  relation: $
 M_P^{1+ \gamma} m_1 = m_2^{2 + \gamma}$ with index $ \gamma $  between
 $0$ in  $+-+$ limit and $1$ in GRS  limit.

\section{Multigravity Zoo}

The idea of multi-localization emerges when one considers
configuration of branes such that the corresponding potential $V(z)$
has at least two ($\delta$-function)\footnote{In the infinitely thin
brane limit that we consider, the wells associated with positive
branes are $\delta$-functions} potential wells . The presence of two
wells that each of them could support a bound state has interesting 
implications. If we consider the above
potential wells separated by an infinite distance, then the zero modes 
are be degenerate i.e. they  have the same mass. However if the
distance between them becomes finite, then as in quantum mechanics,
due to tunneling the degeneracy will be removed and a exponentially
small mass splitting will appear between these states. The rest of 
levels, which are not
bound states, although their mass is  also modified, do not exhibit 
the above exponential splitting. The above becomes even more clear if
one examine the form of the wave-functions. In the finite distance
configuration the wave-function of the zero mode will be the symmetric
combination 
($\hat{f}_{0}=\frac{\hat{f}^{1}_{0}+\hat{f}^{2}_{0}}{\sqrt{2}}$) of
the wave-functions of the zero modes of the two wells (where are
infinitely apart) whereas the wave-function of the first KK state will
be the antisymmetric combination
($\hat{f}_{0}=\frac{\hat{f}^{1}_{0}-\hat{f}^{2}_{0}}{\sqrt{2}}$).  
One can  see that the absolute value of these wave-functions is
almost identical in most of the extra dimension with exception the
central region where the antisymmetric passes through zero thought
the symmetric has suppressed but non-zero value.  Exactly the above is
translated to the exponentially small mass difference of these states.

The phenomenon of multi-localization is of particular interest since
starting from a problem with only one mass scale we are able to create 
a second exponentially smaller scale. Obviously the generation of this
hierarchy is a key to have  
ultra-light massive gravitons, i.e. to have multigravity  
\cite{Kogan:2000wc,Gregory:2000jc,Kogan:2000cv,Mouslopoulos:2000er,Kogan:2000xc}.

\subsection{Universality Classes of Multigravity}
Recently another  other interesting class of 
models with massive gravitons was studied in 
\cite{Dvali:2000hr,Dvali:2001xg} (we shall call it DGP model)
in which they consider a flat bulk 
but took into account an effective action induced on  a brane in 
a one-loop approximation. As been suggested long time ago 
by Sakharov \cite{Sakharov:1968pk} (see also  Zeldovich 
\cite{Zeldovich:1967gd}) D-dimensional  quantum matter induces a 
 a D-dimensional  curvature term (for a review on induced gravity see
 \cite{Adler:1982ri}). 
An effective action  is the sum of four and
five dimensional Einstein-Hilbert actions. Phenomenologically this
model has the same type of behaviour as GRS - crossover from $1/r$
 to $1/r^2$ potential. Actually this is not surprising because  
gravitational low energy effective Lagrangian for GRS   model is 
 precisely  a two-term action of   DGP model.  
The  GRS model in a long wave-length limit looks  like  a sandwich 
with
 positive brane  screened by two negative ones. Effectively this thin
slice of AdS space  contributes an extra term to an effective action
 which plays the  same role  as  induced four-dimensional
 Einstein-Hilbert action - in result one reproduces precisely DGP
effective action. 
However there is one open question - in GRS model there is a
radion field (which we shall discuss in the next section)  which is a
ghost and at first sight there is  no  ghost fields in  DGP model.  in GRS
model the radion field emerges due to fluctuations of $-1/2$ branes - and
of course in a low energy limit these fluctuations are hidden. However it
is not completely clear that  DGP model does not have radion. 
First of all there will be induced
$R^2$  terms and it is well known fact that these terms can  lead to ghost-like
contributions in a graviton propagator. Second problem  is related to the
sign of the induced cosmological constant which must be cancelled by a
brane tension.  If the induced cosmological constant is positive one must
start from the negative brane and it is not clear if there will be no
negative norm states.  However it may happen that  GRS and DGP models are
not equivalent at all. The we have an interesting situation: gravitational
sectors are identical but one model has an extra degree of freedom.

Here we come across a very important concept - {\it Universality Classes
for multigravity}. The concept of universality class is well known  and it
means that theories in the same universality class  may be different at
short scales   but are the same in the infrared limit. For example second
order phase transitions of different physical systems are described by
different universality classes which are completely determined by the
spectrum of anomalous dimensions and symmetries.   It is an interesting
question what are universality classes for multigravity, for example  can
one have two (or more) models with identical  low-energy gravitational
actions but different 
 with  radion/dilaton  low-energy actions ?  If this is impossible GRS and
DGP must be in the same universality class (and  for large scale
phenomenology it does not matter that there may be two completely
different short-distance physics).

Another universality class  was discussed in \cite{Kogan:2001yr}
 where multigravity in six dimensions was  considered. In that case
 first time one got  bounces with flat positive  tension branes. 
Some other models (with AdS branes)  we shall  mention later, but due
to lack of space it is impossible to give any description of these
models here.

Let us note  that because  multigravity is 
intimately  connected with the multilocalization of gravity in 
multibrane constructions \footnote{multilocalization is a property which
can be (with appropriate mass terms) common  for all spins
\cite{Mouslopoulos:2001uc,Kogan:2001wp}} universality classes for
multigravity have a
geometrical interpretation.

How universality classes can be defined ? A full answer on this question
is still unknown but it is clear that the  index  $\gamma$ which  relates
three scales $M_P^{1+ \gamma} m_1 = m_2^{2 + \gamma}$ is a part of a
definition.
 For example $+-+$ and GRS models are in a different universality classes
because they have different $\gamma =0$ or $\gamma = 1$.

\subsection{Can we have massive gravitons ?}
Itis well known that  massive gravitons have extra polarization states 
which do  not decouple in the massless limit - the so-called 
van Dam - Veltman - Zakharov discontinuity 
\cite{vanDam:1970vg,Zakharov}. This property can
 make multigravity phenomenologically unacceptable 
  as was suggested in \cite{Dvali:2000rv}.
However, an equally generic characteristic of some of these  
models (but not all of them !)
 is that they contain moving branes of negative  
tension.  In certain models the radion can help to recover 4D gravity 
on the  
brane at intermediate distances. Indeed, the role of the radion 
associated  
with the negative tension brane is  
precisely to cancel the unwanted  massive graviton polarizations and 
recover  
the correct tensorial structure of the four dimensional graviton 
propagator  
\cite{Dvali:2000km,Pilo:2000et}, something also seen from the bent 
brane calculations of \cite{Csaki:2000ei,Gregory:2000iu}.  
This happens because the radion in this case is a physical ghost 
because
it is has a wrong sign kinetic term. This fact of course makes the 
construction  
problematic because the system is probably quantum mechanically 
unstable. 
Classically,  the origin of the problem  
is the fact that the weaker  energy condition is violated  in the 
presence of moving negative tension branes 
\cite{Freedman:1999gp,Witten:2000zk}.   
 
A way out of this difficulty is to abandon the requirement of flatness  
of the branes and consider curved ones. A particular example was  
provided in the $''++''$ model of \cite{Kogan:2001vb} where no negative 
tension brane was needed to  get multigravity. Moreover, due to the 
fact that the branes where  
$AdS_4$ one could circumvent at least at tree level the van Dam -  
Veltman - Zakharov theorem  \cite{Kogan:2001uy,Porrati:2001cp} and the  
extra polarizations of the massive gravitons where practically  
decoupled. Of course one can ask the question about the resurrection
of these extra polarizations in  quantum loops.  One loop effects in
the massive graviton propagator in $AdS_4$ were discussed  in 
\cite{Dilkes:2001av,Duff:2001zz}. 
Of course, purely four-dimensional theory with massive
graviton is  not  well-defined  and it 
 is certainly true that if the mass term is 
added by hand in purely four-dimensional theory a lot of problems will 
emerge as it was shown in the classical paper of 
\cite{Boulware:1972my}.  If however  the  underlying theory is  a 
higher dimensional one, the graviton(s)
mass terms appear dynamically and this is a different story. 
All quantum corrections must be calculated in a 
higher-dimensional theory, where a larger number of
graviton degrees of freedom is present naturally (a massless
five-dimensional graviton has the same number of  degrees of freedom
as a massive four-dimensional one). 

  Moreover, the smoothness
of the limit $m \rightarrow  0$ 
  is not only a property of the $AdS_4$ space but holds 
for any background where the characteristic curvature invariants are  
non-zero \cite{DamIan,Oxf}. For physical processes taking place 
 in some region a curved space with a characteristic average curvature, 
the effect of graviton mass is controlled by positive powers of the 
ratios  $m^2/R^2$  where $R^2$ is a characteristic curvature invariant
(made
from Riemann and Ricci tensors or scalar curvature).
  A very interesting  argument  supporting  the
conjecture that  there is a smooth limit for phenomenologically 
 observable amplitudes in brane gravity  with ultra-light gravitons 
 is based on a very interesting paper   \cite{Vain}

In that paper it  was shown  that there is a smooth
limit for a   metric  around a spherically symmetric  source with a
mass  $M$   in a theory with
massive graviton  with  mass $m$ for  small (\textit{i.e.} smaller than 
 $m^{-1}(mM/M_P^2)^{1/5}$) distances.
   The discontinuity reveals itself at large distances. The
non-perturbative  solution discussed in \cite{Vain} was found in a 
limited range of distance from
the center and it is still unclear if it can be
smoothly continued to spatial infinity (this problem was stressed in 
 \cite{Boulware:1972my}). Existence of this smooth
continuation depends on the full  nonlinear structure of the
theory. If one adds a mass term by hand the smooth asymptotic at
infinity may not exit.  However, it 
 seems  plausible   that 
 in all cases when modification of gravity at large distances comes
from consistent higher-dimensional models, the global  smooth solution 
can exist because in this case there is 
a unique non-linear structure related to the mass term  which is
dictated by the  underlying 
higher-dimensional theory. In  a  paper \cite{Deffayet:2001uk} an 
example of a 5d cosmological solution  was discussed which contains an
explicit 
interpolation between perturbative and non-perturbative regimes: a 
direct analog
of large and small distances in the Schwarschild case.

 An interesting feature of the above $''++''$ model, which only has a 
dilaton, is that the dilaton survives in the decompactification limit 
when one of the two branes is sent to infinity 
\cite{Papazoglou:2001ed,Chacko:2001em}. This limit was discussed in 
\cite{Karch:2001ct,Miemiec:2000eq,Schwartz:2001ip} and it was found 
that indeed there is a massive scalar mode in the gravity perturbation  
spectrum \cite{Karch:2001ct}. Although it seems strange to 
have a dilaton in an infinite extra dimensional model, it is clear 
that this mode is precisely the remnant of the decompactification 
process of the  compact  $''++''$ model. This happens as we will show 
also to multibrane models with 
flat branes and is related  to the fact that the radion has opposite 
localization properties compared to  the ones of the graviton.  For
 more details about radion dynamics of  multibrane configurations see
\cite{Kogan:2001qx}

The above multibrane constructions in order to be physically acceptable   
should incorporate a mechanism which will stabilize the moduli  
(dilaton, radions) when  
they have positive kinetic energy and will give them some 
phenomenologically  
acceptable mass. This  can be achieved by considering for example a  
bulk scalar field 
\cite{Goldberger:1999uk,DeWolfe:2000cp,Kanti:2000rd,Csaki:2001zn} with 
non
trivial bulk  
potential (for the effect of the Casimir force between the branes see 
\cite{Hofmann:2000cj}). A general condition that guarantees  
stabilization of the dilaton in the case of maximally symmetric branes  
was derived in \cite{Papazoglou:2001ed} and restricts the sum of the 
effective  
tensions of the branes and the leftover curvature of the brane.  The  
moduli stabilization has greater importance in the context of brane  
cosmology where it was found that it played a crucial role in deriving   
normal cosmological evolution on the branes  
\cite{Kanti:1999sz,Csaki:2000mp,Kanti:2000nz}.  A  
non-perturbative analysis of the dilaton  two  
brane models can be found in \cite{Binetruy:2001tc}. 
  
 When the radion has negative 
kinetic energy is still not clear whether one can speak about 
stabilization of these systems. They are probably unstable 
at the quantum level and no one has attempted to 
estimate their life-time. Actually it is not completely clear
 if unstable negative norm states are really  forbidden - the problem
is that if there are several excitations in the same channel and they
all have finite width an interesting thing may happen - the
distributions with negative norms will sit inside positive norm
resonances. In other words total spectral density may be still
positive - unstable negative norm states will simply reduce total
spectral density.  Perhaps one can think that unwanted polarizations
of massive gravitons in physical amplitudes will be cancelled by
negative norm states which becomes resonances and in result one will
have totally positive spectral density. This is still an open question 
which deserves a further investigation.

\section{Observable effects}

Observation of modifications of gravity at ultra-large scale could be a   
striking signal of such a possibility. There are several papers in which
experimental tests were discussed.  In \cite{Binetruy:2001xv} CMB 
 measurements were  discussed which are sensitive to a large scale 
 limit of gravitational interactions.   
Lensing at cosmological scales due to multigravity 
was discussed in \cite{Uzan:2000mz} and SN1A data and CMB  for
multigravity 
 was considered in  \cite{Bastero-Gil:2001rv}. A lot of astrophysical
constraints on modifying gravity at large distances
 was  discussed in \cite{Aguirre:2001xs}. One of the reason why
multigravity
  can modify CMB is that it leads to a 
large distance modifications of the curvature. When traced back at the
time of inflation this gives rise to a dispersed frequency for the cosmic
perturbations. The perturbation field is minimally coupled to gravity 
 and it long wavelength modes   are influenced by  the
modifications in the background curvature. The analysis of the CMB
spectrum for the whole range of modes \cite{Bastero-Gil:2001rv} reveals
 that the spectrum deviates from  scale-invariance and is extremely
sensitive to 
large distance physics (because long wavelength modes dominate
the spectrum) and the choice of the initial conditions, but does not depend 
in the details of short-distance physics (transplanckian modes).This
deviation is small for curvature modifications around the last 
scattering horizon scale. 

There are other experiments  to study  multigravity, for example
precision tests, but here we do not have time to discuss it.

\subsection{Multigravity and Cosmology:Dark Matter and Quintessence}

One of  very striking features of multigravity is that it gives us a
some  sort of a dark matter.  The origin of this dark matter is very
straightforward - this is just matter from other branes! So it is by
definition "dark" if we assume that all SM fields propagates along the
same brane.  But why shall we see matter from other branes ?  Let's
concentrate on 5-dimensional case.
We have multilocalization    and the massless graviton is localized on
several branes - which means that it interaction with matter is of the
form:
 \bea
 h^{0}_{\mu\nu}(x) \sum_{i} \Psi_{0}(y_i) T_{\mu\nu}^{i}(x)
\eea
where $ T_{\mu\nu}^{i}(x)$ are stress-energy tensors of matter localized
on  different branes  located at  points $y_i$ and $\Psi_{0}(y)$ is the
massless  graviton profile (which must be of the bounce form). 
The coupling for the next excited state is given by 
 \bea
 h^{1}_{\mu\nu}(x) \sum_{i} \Psi_{1}(y_i) T_{\mu\nu}^{i}(x)
\eea
where  $\Psi_{1}(y)$ is the first ultra-light graviton profile, etc..
Because $ \Psi_{n}(y_i)$  are quite different we see that we have
different effective source. Let us study the simplest model with two
branes, when on a first  brane 
$\Psi_{0}(y_1) = \Psi_{1}(y_1) = 1$ and on  a second one
$\Psi_{0}(y_2) = -\Psi_{1}(y_2) = 1$
the coupling is
\bea
\left(h^{0}_{\mu\nu}(x)+h^{1}_{\mu\nu}(x)\right)T_{\mu\nu}^{1}(x)
+
\left(h^{0}_{\mu\nu}(x)-h^{1}_{\mu\nu}(x)\right)T_{\mu\nu}^{2}(x)
\eea
On a first brane observable metric is $
h^{0}_{\mu\nu}(x)+h^{1}_{\mu\nu}(x)$ and we see that matter from
second brane does not contribute to it. But this is true only
 for distances  smaller than $R_{C}$ when we can not see difference
between two gravitons. For larger distances the ultra-light graviton
effectively decouples and we can only see the massless one. But
 $h^{0}_{\mu\nu}(x)$ is produced by the {\it sum} of stress-energy
tensors from both branes $T_{\mu\nu}^{1}(x) + T_{\mu\nu}^{2}(x)$.
 Thus multigravity opens a  window in extra dimensions - we start to
feel gravitationally matter which is localized  on other branes !
 Moreover - we can not see this matter at intermediate scales - the
coherent combination of massless and light gravitons screen this
matter from us. This window opens  ONLY at  large scale.

 This is indeed a dark matter. In an example described above we shall
see two times more matter -but at the same time our gravitational
constant  is smaller by a factor of 2 so overall gravity is not
stronger. But  we can consider more sophisticated models with more
than 2 branes and in principle it is possible to get an amplification
of the gravitational force. 
It  remains to be seen if  one can construct a  model which will be
phenomenologically acceptable to explain
rotational curves. 

But in any case we see that multigravity means dark matter, the
question is can it account for ALL dark matter ?
This question is  closely related to another one:
 what effect does  have  multigravity on
 cosmology ? The Hubble law relates rate of expansion to matter
density $(\dot{a}/a)^2 \sim  G_N \rho$.
 Change in $G_N$ leads to change in Hubble law. This means that for
scale factors $a(t)$ bigger and smaller the crossover radius $R_C$
 we have to use different gravitational constants. Thus crossover in
Newton law  leads to a crossover in cosmological evolution. Does it
mean that multigravity also means Quintessence ? Is multigravity
important for cosmological constant problem ?

\section{Conclusion}
It seems it is a good time to stop here.  Answers on all these
exciting questions are not known (or known only partially).  
This is  what people are working on 
right now and hopefully one day we shall know these answers.  If they are
negative - well, this is just another interesting possibility which
has nothing to do with real world (or we have some small admixture of
ultralight states, but nothing drastic). But if the answers are
positive.....

\section*{Acknowledgments}
This talk was  given at Moriond meeting and  memories of this exciting 
(as usually) event helped a lot during the process of writing. I am
grateful for all organizers of this wonderful meeting for very
stimulating and relaxed atmosphere. Special thanks  are  to 
Claude Barthélemy - who was persistent enough to persuade me to write
this text  after two deadlines and patient enough for me not to feel 
guilty. I am indebted  to  Stavros Mouslopoulos, Antonios Papazoglou and 
Graham G. Ross  in collaboration with whom most of
the results presented here were obtained.  I  would like to thank  
Thibault Damour, Panagiota Kanti, Laura Mersini, Keith Olive, Luigi
Pilo, Bayram Tekin and Arkady Vainshtein  for  very
stimulating discussions. This work   is  
supported in part by the PPARC rolling grant PPA/G/O/1998/00567, by 
the EC TMR grants  HRRN-CT-2000-00148 and  HPRN-CT-2000-00152.

\end{document}